\documentclass[10pt,twocolumn,twoside]{IEEEtran}

\usepackage{cite}
\usepackage{amsmath,amssymb,amsfonts}
\usepackage{algorithmic}
\usepackage{graphicx}
\usepackage{textcomp}

\usepackage{mathdots}
\def\BibTeX{{\rm B\kern-.05em{\sc i\kern-.025em b}\kern-.08em
    T\kern-.1667em\lower.7ex\hbox{E}\kern-.125emX}}

\usepackage[table]{xcolor}    
\usepackage{array}  
\usepackage{cancel} 

\usepackage{subfig}

\newtheorem{theorem}{Theorem}

\newtheorem{remark}{Remark}

\newtheorem{definition}{Definition}
\newtheorem{assumption}{Assumption}

\DeclareMathOperator*{\diag}{diag}

\begin{document}

\title{From classical to Networked Control: \\Retrofitting the Concept of Smith Predictors
}
\author{Martin Steinberger, Martin Horn
	
\thanks{M. Steinberger and M. Horn were supported by the LEAD project ``Dependable Internet of Things in Adverse Environments'' funded by Graz University of Technology. The financial support by the Christian Doppler Research Association, the Austrian Federal Ministry for Digital and Economic Affairs and the National Foundation for Research, Technology and Development is gratefully acknowledged. 
}
\thanks{M. Steinberger and M. Horn are with the Institute of Automation and Control, Graz University of Technology, 8010 Graz, Austria (e-mail: martin.steinberger@tugraz.at; martin.horn@tugraz.at).}
\thanks{M. Horn is with the Christian Doppler Laboratory for Model Based Control of Complex Test Bed Systems, Institute of Automation and Control, Graz University of Technology, 8010 Graz, Austria.}
}

\maketitle

\begin{abstract}
	Filtered Smith predictors are well established for controlling linear plants with constant time delays. Apart from this classical application scenario, they are also employed within networked control loops, where the measurements are sent in separate packets over a transmission channel that is subject to time-varying delays. However, no stability guarantees can be given in this case.	
	The present paper illustrates that the time-varying delays as well as the packetized character of the transmissions have to be taken into account for stability analysis. Hence, three network protocols, which use different packet selection and hold mechanisms, are considered. Criteria for robust stability of the networked feedback loop are given. They are based on the small gain theorem and allow a computationally inexpensive way to check stability for the case with bounded packet delays. Simulation examples provide insight into the presented approach and show why the inclusion of the time-varying packetized character of the network transmissions is vital for stability analysis.
\end{abstract}


\section{Introduction}

In the field of networked control, significant effort was taken to master networked induced effects like time-varying transmission delays and packet dropouts \cite{Gupta2010, Hespanha2007}. This yields advances in terms of controller design, stability analysis and wireless network design, see, e.g., \cite{Zhang2017}, \cite{Park2018} and the references therein.
Depending on the assumptions about the network connection, stochastic approaches as, e.g., in \cite{Schenato2007}, \cite{Palmisano2021}, or methods only assuming the boundedness of delays are employed \cite{Heemels2010}. For example, predictive controllers are used in \cite{Liu2010} and \cite{Mu2005}. Controller design and stability analysis based on over-approximation techniques and linear matrix inequalities (LMIs) are presented in \cite{Cloosterman2010} and \cite{Heemels2010}. Gain scheduling state feedback integral controllers are designed in 
\cite{Li2011b}. In \cite{Ludwiger2018}, \cite{Ludwiger2019}, combinations of a buffering mechanism together with different sliding mode techniques are used to robustly stabilize networked control systems. Smart actuators and a combination of model predictive control and integral sliding mode control are proposed in \cite{Incremona2017} to reduce the resulting network traffic. 

A classical tool to take into account delays in feedback loops was originally introduced by O. Smith \cite{Smith1959}, who presented the concept for stable plants with constant time delays.
Several extension of the original Smith predictor can be found in literature. In \cite{Normey2009}, a version of a filtered Smith predictor is proposed that allows to robustly stabilize uncertain unstable plants with constant time delays. This work is extended to multivariable systems with different time delays, see \cite{Santos2016}.

Recently, this approach was utilized also in the context of networked control. 
In \cite{Lai2010}, the original Smith predictor in combination with a PI-controller is used for a networked loop using Ethernet and CAN. The round-trip time, i.e. the time for a packet on the way between the transmitter at the sensor side and the receiver at the actuator side, is measured and directly used as nominal delay in the Smith predictor.
The authors of \cite{Repele2014} use the mean value of the estimated actual delay together with an adaptive buffer resizing mechanism in the Smith predictor with PI-controller.
A Smith predictor using the average delay for successfully transmitted packets is utilized in \cite{Gamal2016} for IEEE 802.15.4 wireless personal area networks. A similar approach was followed in \cite{Batista2018} for an optical oven that is connected via CAN and the internet to the controller. Again, the moving average of measured delays is used in the Smith predictor. 
The common denominator of the above mentioned papers using standard Smith predictors for networked feedback loops is that, although time-varying delays are present, only constant or averaged values for the time delays are considered. In addition, no stability analyses respecting the time-varying character of the feedback loops are conducted.
 
Alternative ways are followed, e.g., in \cite{Bonala2017} where polytopic over-approximation and LMI conditions are used to show stability of the standard Smith predictor within a networked loop. In \cite{Normey2012}, a robust stability analysis of the filtered Smith predictor \cite{Normey2009} for time-varying delay processes is introduced. It makes use of delay dependent LMI conditions to show stability of the closed loop, where the uncertainties in the plant are assumed to be norm-bounded. However, the packetized character of the data transmission is not taken into account in this paper. The present paper aims to close this gap.

In real communication channels, each transmitted packet is associated with a corresponding network delay, e.g., due to the actual network load. It is a matter of fact that this packetized character is neglected in most works in literature. On one hand, this missing network property yields incorrect simulation results as shown in detail in \cite{Steinberger2020}, where simulation approaches using Matlab standard blocks \cite{Mathworks2012} and the toolbox TrueTime \cite{Cervin2003} are investigated and extended such that the packetized nature of the transmissions is taken into account. On the other hand, stability analysis, e.g., based on LMI conditions as in \cite{Li2011}, \cite{Seuret2015}, might be problematic due to the neglected packetized character. One possibility to circumvent the effect of potential packet disordering, e.g., in wireless multi-hop networks, is shown in \cite{Liu2015} and \cite{Wu2018}, where reordering mechanisms are introduced to change the underlying network policy.

As a basis for the present work, the small gain \cite{Kao2004}, \cite{Sastry1999} based stability criterion for networked feedback loops with bounded time-varying delays presented in \cite{Steinberger_2020_LCSS} is used. The present work extends this approach with respect to several aspects. It exhibits the following main contributions:
\begin{enumerate}
	\item[(a)] It is shown that stability conditions which do not explicitly account for the packetized character of the network are problematic. This is demonstrated for a LMI-based approach by means of a simulation example.
	
	\item[(b)] An easy to use stability criterion for the filtered Smith predictor with a linear plant and a packetized transmission network is proposed. Time delays are assumed to be bounded and time-varying.

	\item[(c)] The effects of three different network protocols that differ in the packet skipping and hold mechanisms are investigated. It is proven that, e.g., the maximal admissible network delay is lower for the case where no numbering of the transmitted packets is employed. 

	\item[(d)] Finally, an extension of the stability analysis for the networked filtered Smith predictor in closed loop with uncertain plant models is proposed.
\end{enumerate}
The paper is structured as follows: Section~\ref{sec:problem} describes the problem formulation and introduces the considered protocols.  Section~\ref{sec:LMI} presents one possible way to check stability using LMIs and why this might fail. Consequently, Section~\ref{sec:SGT} introduce new stability criteria using nominal and uncertain plant models respectively. Conclusion are drawn in Section~\ref{sec:conclusion}.\medskip

\emph{Notation:} A discrete-time transfer function is written as $G(z)$. 
With the term magnitude plot of a $G(z)$ we mean the magnitude plot of the corresponding frequency response for $z = e^{j \omega h}$ and frequencies $\omega\in \left[0,\pi/h\right)$. Constant $h$ represents the sampling time. The infinity norm $\vert\vert G(z)\vert\vert_{\infty}$ equals the maximum of the magnitude plot of $G(z)$. Entire sequences are denoted by $(y_k) = (y_0, y_1, y_2, \ldots)$, one element is written as $y_k$, where $k\in\mathbb{N}$ is the iteration index. The 2-norm of sequence $(y_k)$ is expressed as $\big|\big|(y_k)\big|\big|_2$. Operator $\mathcal{Z}\left\{(y_k)\right\}$ represents to the z-transform of sequence $(y_k)$. Matrix $I_a$ is the identity matrix with $a$ rows and columns, $0_{a\times b}$ stands for a zero matrix with $a$ rows and $b$ columns.

\section{Problem Formulation}
\label{sec:problem}

Consider a linear time-invariant discrete-time plant model $P(z)$ consisting of a nominal, delay-free part $\hat P(z)$, a nominal constant time delay $\hat d$ and a multiplicative uncertainty represented by $\delta P$ such that
\begin{equation}\label{eq:Pz}
P(z) = \hat P(z) z^{-\hat d} \big(1+\delta P\big) \, .
\end{equation}
A filtered Smith predictor structure \cite{Normey2009} is utilized to allow to control this (possibly unstable) plant with time delay. It consists of controller $C(z)$ and prefilter $V(z)$ designed for the delay-free plant. In addition, two transfer functions $F(z)$ and $H(z)$ are employed in the filtered Smith predictor as described below.
\begin{figure}[!t]\centering
	\includegraphics[scale=0.4]{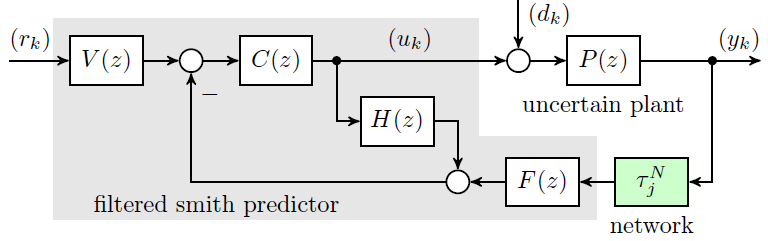}
	\caption{Structure of the closed loop comprising an uncertain plant and a filtered Smith predictor that is closed via a network connection, which is subject to time-varying delays $\tau_j^N$ for individual packets $j$.}
	\label{fig:filtered_smith_nw}
\end{figure}
Figure~\ref{fig:filtered_smith_nw} depicts the structure of the feedback loop, where sequence $(r_k)$ is the reference signal and $(y_k)$ symbolizes the output signal that is sent over a communication network (green block). Sequence $(w_k)$ stands for external disturbances acting on the plant.
\begin{assumption}[Network delay]\label{as:delay}
	The output signal is sent in separate packets $j$ that experience an individual bounded packet delay
	\begin{equation}\label{eq:NW_delay}
		0 \le \underline\tau^N \le \tau_j^N \le \bar\tau^N 
	\end{equation}
	with $\underline\tau^N, \tau_j^N, \bar\tau^N\in \mathbb{N}$, and $\bar\tau^N> 0$.	
\end{assumption}
\begin{remark}
	We distinguish between packet index $j$ and iteration index $k$ to be able to link delays to packets. At one sampling instant $k$ there might be more arriving packets $j$ at the receiver side that experienced different delays $\tau_j^N$.
\end{remark}

The main point to consider for the networked connection is, apart from of the bounded delay according to Assumption~\ref{as:delay}, the packetized character of the transmissions of output sequence $(y_k)$. Thus, the actual realization of a network policy (protocol) has to be taken into account. Three different protocols are formally introduced in the following. A packet containing output sample $y_k$ is sent at instant $k$. It is received at the controller at sampling instant $r^{(k)}\in\{k, k+1, \ldots, k+\bar\tau-1,k+\bar\tau\}$ depending on the actual packet delay $\tau_j$. Since the maximal packet delay is bounded, the packet containing $y_k$ reaches the receiver at the controller side at the latest at $r^{(k)}=k+\bar\tau$. The packet including $y_{k+1}$ arrives at $r^{(k+1)}\in\{k+1, k+2,\ldots, k+\bar\tau,k+\bar\tau+1\}$. Consequently, the arrival instant of the packet containing $y_{k+\bar\tau}$ equals  $r^{(k+\bar\tau)}\in\{k+\bar\tau, k+\bar\tau+1,\ldots, k+2 \bar\tau -1,k+2\bar\tau\}$. At instant $p=k+\bar\tau\in\mathbb{N}$, at most $\bar\tau+1$ packets may arrive that contain $y_k$ if $r^{(k)}=p$, $y_{k+1}$ if $r^{(k+1)}=p$, $\ldots$, $y_{k+\bar\tau}$ if $r^{(k+\bar\tau)}=p$. Hence, the set of all received output samples at instant $p$ is
$\mathcal{Y}(p) = \{y_{k+i} \ \vert \ r^{(k+i)}=p, \ i=0,1,\ldots,\bar\tau\}$ and the set of all indices of received output samples is
\begin{equation} 
	\mathcal{I}(p) = \Big\{ \kappa \ \big\vert \ r^{(\kappa)}=p, \ \kappa=k,k+1,\ldots,k+\bar\tau \Big\} \, .
\end{equation}
\begin{definition}[Protocols] \label{def:protocols}
	The receiver at the controller side chooses a plant output sample $\hat y_p$ at time instant $p\in\mathbb{N}$ depending on the protocol such that for
	\begin{subequations}\label{eq:prot}
	\begin{enumerate}
		\item[(i)] protocol $\mathcal{P}_1$:
			\begin{equation}\label{eq:p_skip}
				\hat y_p = 
				\begin{cases}
					y_{\bar\kappa_p},\ \bar\kappa_p = {\displaystyle\max_{\kappa\in\mathcal{I}(p)}} \kappa  & \text{if}\ \mathcal{I}(p)\neq\{\}\ \wedge \\
					 & \bar\kappa_p > \bar\kappa_{p-1}  \\
					\hat y_{p-1}     & \text{otherwise}
				\end{cases}
			\end{equation}
		
		\item[(ii)] protocol $\mathcal{P}_2$:
			\begin{equation}\label{eq:p_noskip}
				\hat y_p = 
			\begin{cases}
				y_{\bar\kappa_p},\ \bar\kappa_p = {\displaystyle\max_{\kappa\in\mathcal{I}(p)}} \kappa & \text{if}\ \mathcal{I}(p)\neq \{\}  \\
				\hat y_{p-1}     & \text{otherwise}
			\end{cases}
			\end{equation}
		
		\item[(iii)] protocol $\mathcal{P}_3$:
			\begin{equation}\label{eq:p_worst}
				\hat y_p = 
				\begin{cases}
					y_{\bar\kappa_p},\ \bar\kappa_p=\text{any}(\kappa)\in\mathcal{I}(p)  & \text{if}\ \mathcal{I}(p)\neq \{\}  \\
					\hat y_{p-1}     & \text{otherwise}
			\end{cases}
			\end{equation}
	\end{enumerate}
	\end{subequations}
\end{definition}
\begin{remark}
	The cases using $\hat y_{p-1}$ in \eqref{eq:p_skip}-\eqref{eq:p_worst} constitute a zero order hold mechanism for each packet that is active whenever no packet is received at the actual time instant $p$, i.e. $\mathcal{I}(p)= \{\}$.
	 The protocols differ in terms of the packet selection mechanisms. In protocol $\mathcal{P}_2$, the newest available packet with the corresponding index $\bar\kappa_p$ is selected. In addition to that, $y_{\bar\kappa_p}$ is only employed according to protocol $\mathcal{P}_1$ if the actual available packet at time instant $p$ is newer than the packet used at the previous time instant $\bar\kappa_{p-1}$, i.e. $\bar\kappa_p > \bar\kappa_{p-1}$. As a consequence, packets that are older than the previously selected packet are skipped. Note that a unique packet number has to be attached to each sent packet but no synchronization, e.g., using \cite{IEEE2008}, is needed between the transmitter at the sensor side and the receiver at the controller side. 
	 In situations where no packet numbering is implemented for simplicity reasons, protocol $\mathcal{P}_3$ is applied. It selects any of the actual received packets at time instant $p$, e.g., based on the position in the internal packet buffer. 
\end{remark}

The output of the network block in Fig.~\ref{fig:filtered_smith_nw} is fed into transfer function $F(z)$ of the filtered Smith predictor. The originally proposed version of the predictor \cite{Normey2009} is designed for a plant with constant time delay. Hence,
the overall time delay of the plant and the communication network is split into two parts. Transfer function 
\begin{equation}\label{eq:Dz}
	\hat D(z) = z^{-\hat d -\underline\tau^N} = z^{-\hat\tau}
\end{equation}
combines the constant plant delay $\hat d$ and the minimum delay $\underline\tau^N$ introduced by the network. The remaining variable time delay $\tau_j$ is then bounded by
\begin{equation}\label{eq:tau_bounds}
	0 \le \tau_j \le \bar\tau^N-\underline\tau^N = \bar\tau \, .
\end{equation}

The three design steps for the networked filtered Smith predictor shown in Fig.~\ref{fig:filtered_smith_nw} are as follows:

(Step~1) Design of a nominal controller $C(z)$ for the nominal plant $\hat P(z)$.
	
(Step~2) Design of a stable filter transfer function $F(z)$ with a dc-gain equal to one, i.e.
\begin{equation}\label{eq:Fz}
	F(z) = \frac{\mu_F(z)}{\nu_F(z)} \quad\text{with}\quad  F(1)=1 \, ,
\end{equation}
and transfer function $H(z)$ for the overall minimum delay $\hat\tau$ as stated in \eqref{eq:Dz}, such that 
\begin{equation}\label{eq:Hz}
	H(z) = \hat P(z) \left(1-z^{-\hat\tau} F(z)\right) = \frac{\hat\mu(z) \left(z^{\hat\tau} \nu_F(z) - \mu_F(z)\right)}{z^{\hat\tau} \hat\nu(z) \nu_F(z)} 
\end{equation}
holds with $\hat P(z) = \frac{\hat\mu(z)}{\hat\nu(z)}$.  
Since the denominator polynomial $\hat\nu(z)$ of the nominal plant might have unstable zeros, one has to choose 
\begin{equation}\label{eq:design_Hz}
	z^{\hat\tau} \nu_F(z) - \mu_F(z) = 0 \ , \quad \forall z \ \big\vert \ \hat\nu(z)=0 \ \wedge\ |z|\ge 1 
\end{equation}
to obtain a stable $H(z)$. The poles of $F(z)$, i.e. the zeros of polynomial $\nu_F(z)$ could, for example, be selected to be at the same positions $z=\lambda$, where $0\le \lambda < 1$. It is also possible to modify \eqref{eq:design_Hz} to compensate for stable but ``slow'' poles in $\hat P(z)$. Parameter $\lambda$ allows to tune the conflicting properties of disturbance rejection and robustness with respect to plant uncertainties, see \cite{Normey2009} for details.

For the nominal case, i.e. without uncertainties $\delta P$ and constant time delay $\bar\tau=0$ \eqref{eq:tau_bounds}, the closed loop transfer functions relating external inputs and the plant output can be composed using
\begin{align}
	\hat T_{r}(z) &= \frac{\mathcal{Z}\big\{(y_k)\big\}}{\mathcal{Z}\big\{(r_k)\big\}}  =  \frac{V(z) C(z) \hat P(z) \hat D(z)}{1+C(z) \hat P(z)} \, ,\\
	\hat T_{d}(z) &= \frac{\mathcal{Z}\big\{(y_k)\big\}}{\mathcal{Z}\big\{(d_k)\big\}}  = \hat P(z) \hat D(z) \Bigg[ 1- \frac{C(z) F(z) \hat P(z) \hat D(z)}{1+C(z) \hat P(z)} \Bigg] \nonumber\, ,
\end{align}
\eqref{eq:Dz} and \eqref{eq:design_Hz}. Relation \eqref{eq:design_Hz} yields stable predictions of the output using $H(z)$ and, as a consequence, an internally stable feedback loop for the nominal case \cite{Normey2009}.

(Step~3) In the final design step, one has to show that the filtered Smith predictor designed in Step~2 also leads to a stable feedback loop for the case with bounded time-varying packet delays ($\bar\tau>0$), different transmission protocols \eqref{eq:prot}, and uncertain plants ($\delta P\neq 0$). This challenging task can be solved by means of the stability criteria proposed in the subsequent sections.

\section{Stability Analysis based on Linear Matrix Inequalities}
\label{sec:LMI}

In this section, a theorem to check the stability of the feedback loop in Fig.~\ref{fig:filtered_smith_nw} for the case without plant uncertainties, i.e. $\delta P= 0$, is formulated.

\subsection{LMI-based stability criterion}

It is based on LMI conditions proposed in \cite{Li2011}. Hence, all transfer functions shown in Fig.~\ref{fig:filtered_smith_nw} are written as minimal realizations such that
\begin{subequations}\label{eq:state_space_models}
	\begin{eqnarray}
	&P(z)& \begin{cases}
	x_{k+1} = A x_k + b u_k \\
	\quad\, y_k = c^T x_{k-\hat d}
	\end{cases}\label{eq:state_space_model_P}\\
	&H(z)& \begin{cases}
	x_{H,k+1} = A_H x_{H,k} + b_H u_k \\
	\quad\, y_{H,k}   = c_H^T x_{H,k}
	\end{cases}\label{eq:state_space_model_H} \\
	&F(z)& \begin{cases}
	x_{F,k+1} = A_F x_{F,k} + b_F y_{k-\tau_k^N} \\
	\quad\, y_{F,k}   = c_F^T x_{F,k} + d_F y_{k-\tau_k^N}
	\end{cases} \label{eq:state_space_model_F}\\
	&C(z)& \begin{cases}
	x_{C,k+1} = A_C x_{C,k} - b_C \left(y_{H,k}+y_{F,k}\right) \\
	\quad\, y_{C,k}   = c_C^T x_{C,k} - d_C \left(y_{H,k}+y_{F,k}\right)
	\end{cases} \label{eq:state_space_model_C}
	\end{eqnarray}
\end{subequations}
with state vectors $x_k\in\mathbb{R}^n$, $x_{H,k}\in\mathbb{R}^{n_H}$, $x_{F,k}\in\mathbb{R}^{n_F}$ and $x_{C,k}\in\mathbb{R}^{n_C}$. Note that the constant nominal plant delay $\hat d$ and the time-varying network delay $\tau_k^N$ are present in \eqref{eq:state_space_model_P} and \eqref{eq:state_space_model_F}, respectively. The combination of all sub-models \eqref{eq:state_space_models} according to Fig.~\ref{fig:filtered_smith_nw} yields the closed loop description
\begin{equation}\label{eq:state_space_model}
	\xi_{k+1} = \tilde A \xi_k + \tilde A_d \xi_{k-\hat d-\tau_k^N}
\end{equation}
with state vector
\begin{equation}\label{eq:state_vector}
	\xi_k^T = \begin{bmatrix}
	x_k^T & x_{H,k}^T & x_{F,k}^T & x_{C,k}^T
	\end{bmatrix} \in\mathbb{R}^{n_\xi}
\end{equation}
and
\begin{subequations}\label{eq:state_space_matrices}
\begin{align}
	\tilde A &= \begin{bmatrix}
		A & -b\,d_C\,c_H^T & -b\,d_C\,c_F^T & b\,c_C^T\\
		0 & A_H-b_H\,d_C\,c_H^T & -b_H\,d_C\,c_F^T & b_H\,c_C^T\\
		0 & 0 & A_F & 0\\
		0 & -b_C\,c_H^T	& -b_C\,c_F^T & A_C 
	\end{bmatrix}\, ,\\
	\tilde A_d &= \begin{bmatrix}
		-b\,d_C\,d_F\,c^T	  & 0 & 0 & 0 \\
		-b_H\,d_C\,d_F\,c^T & 0 & 0 & 0  \\
		b_F\,c^T          & 0 & 0 & 0  \\
		-b_C\,d_F\,c^T     & 0 & 0 & 0  \\			
	\end{bmatrix} \, .
\end{align}
\end{subequations}
Based on that, a stability criterion for the networked filtered Smith predictor in closed loop with a known plant \eqref{eq:Pz} with $\delta P =0$ is formulated.
\begin{theorem}[LMI stability criterion]\label{th:stability_LMI}
	Consider the networked filtered Smith predictor as shown in Fig.~\ref{fig:filtered_smith_nw}, where transfer functions $V(z)$ and $C(z)$ are designed to stabilize the nominal, delay-free Plant $\hat P(z)$ and $\hat\tau = \hat d +\underline\tau^N >0$ \eqref{eq:Dz}. The plant \eqref{eq:Pz} is assumed to be known, i.e. $\delta P =0$. Transfer functions $F(z)$ and $H(z)$ are selected according to \eqref{eq:Fz}, and \eqref{eq:Hz}, \eqref{eq:design_Hz}, respectively. 
	
	Then, the closed loop is asymptotically stable for all bounded time-varying network delays $\tau_k^N$ \eqref{eq:NW_delay} if one of the following conditions holds for a scalar constant $0 < \gamma < 1$:
	\begin{enumerate}
		\item[(i)] There exist positive definite symmetric matrices $P\in\mathbb{R}^{\left(\hat d + \bar \tau^N+1\right) n_\xi\times \left(\hat d + \bar\tau^N + 1\right) n_\xi}$ and $S\in\mathbb{R}^{n_\xi \times n_\xi}$ so that
		\begin{equation}
			G^T \Theta_1 G - \Theta_2 \prec 0
		\end{equation}
		with matrices $\Theta_1 = \diag{(P,S)}$, $\Theta_2 = \diag{(P,\gamma^2 S)}$ and
		\begin{equation}
			G = \begin{bmatrix}
				\Psi_1 & \frac{1}{2} \tilde A_d & \frac{\bar\tau^N - \underline\tau^N}{2} \tilde A_d\\[3pt]
				I_{\left(\hat d + \bar \tau^N\right) n_\xi} & \left(0\right)_3 & \left(0\right)_3\\[3pt]
				\Psi_2 & \frac{1}{2} \tilde A_d & \frac{\bar\tau^N - \underline\tau^N}{2} \tilde A_d
			\end{bmatrix}
		\end{equation}
		where
		\begin{align}
			\Psi_1 &= \begin{bmatrix}
				\tilde A & \left(0\right)_1 & \frac{1}{2} \tilde A_d & \left(0\right)_2
				\end{bmatrix}\nonumber\\
			\Psi_2 &= \begin{bmatrix}
				\tilde A-I_{n_\xi} & \left(0\right)_1 & \frac{1}{2} \tilde A_d & \left(0\right)_2
				\end{bmatrix}	\\
			\left(0\right)_1 &= 0_{n_\xi\times \left(\hat d + \underline\tau^N 
					-1\right)n_\xi} \, ,\quad
			\left(0\right)_2 = 0_{n_\xi\times \left(\bar\tau^N - \underline\tau^N 
					- 1 \right)n_\xi} \nonumber\\
			\left(0\right)_3 &= 0_{\left( \hat d + \bar\tau^N \right) n_\xi\times  n_\xi}\nonumber
		\end{align}
		
		\item[(ii)] There exist positive definite matrices $P, Q_1, Q_2, R_1, R_2,$ $S \in\mathbb{R}^{n_\xi\times n_\xi}$ so that
		\begin{equation}
			\begin{bmatrix}
			\Phi_1 & \Psi\\
			\ast & \diag{\left(-P,-R_1,-R_2,-S\right)}
			\end{bmatrix} \prec 0
		\end{equation}
		where
		\begin{align}
			\Phi_1 &= \begin{bmatrix}
			\Phi_{11} & \begin{bmatrix}
			R_1 & R_2 & 0_{n_\xi} 
			\end{bmatrix} \nonumber\\[3pt]
			\ast & \diag{\left(\Phi_{12},\Phi_{13},-\gamma^2 S\right)}
			\end{bmatrix}\nonumber\\
			\Phi_{11} &= -P + Q_1 + Q_2 - R_1 -R_2\\
			\Phi_{12} &= -Q_1 - R_1 \, , \quad
			\Phi_{13} = -Q_2 - R_2 \nonumber\\
			\Psi &= \footnotesize\begin{bmatrix}
			\Phi_2^T P & \left(\hat d + \underline \tau^N\right)\Phi_3^T R_1 & \left(\hat d + \bar\tau^N\right)\Phi_3^T R_2 & \Phi_3^T S
			\end{bmatrix}\nonumber\\			
			\Phi_2 &= \begin{bmatrix}
			\tilde A & \frac{1}{2} \tilde A_d & \frac{1}{2} \tilde A_d & \frac{\bar\tau^N - \underline\tau^N}{2} \tilde A_d
			\end{bmatrix} \nonumber\\
			\Phi_3 &= \begin{bmatrix}
			\tilde A-I_{n_\xi} & \frac{1}{2} \tilde A_d & \frac{1}{2} \tilde A_d & \frac{\bar\tau^N - \underline\tau^N}{2} \tilde A_d
			\end{bmatrix}\quad	\nonumber	
		\end{align}

	\end{enumerate}	
\end{theorem}

\begin{IEEEproof}
	The proof is a direct consequence of Corollary~1 in \cite{Li2011} evaluated for \eqref{eq:state_space_model}, \eqref{eq:state_space_matrices}, $h_1 = \hat\tau = \hat d +\underline\tau^N$, $h_2=\hat d +\bar\tau^N$, and $h_{12}=\bar\tau^N - \underline\tau^N$.
\end{IEEEproof}

\begin{remark}
	Note that Theorem~\ref{th:stability_LMI} differs from the stability criterion presented in \cite{Normey2012} in several ways. The main difference is that the time varying-delay is modeled in the input channel of the plant in \cite{Normey2012}. It is not equivalent the situation presented in Fig.~\ref{fig:filtered_smith_nw}, where the output sequence is transmitted via a time-varying network channel. In addition, a different Lyapunov-Krasovskii functional is employed in the proof to get the LMI conditions.
\end{remark}

\subsection{Simulation example: LMI-based approach}
\label{sec:example1}

A simple simulation example is shown in this section to underline the achieved properties. The plant is given by
\begin{equation} \label{eq:ex_plant}
	P(s) = \frac{0.1}{20 s -1} e^{-5 s} \quad\text{and}\quad P(z) = \frac{0.0051271}{z-1.051} z^{-5} \, ,
\end{equation}
respectively, where $h=1\, s$ was selected as sampling time. A nominal controller $C(z)$ was designed assigning the poles of the nominal, delay-free, closed loop to be at $p_1 = p_2 = 0.95$. Additionally, the controller should include an integral action to allow to compensate for constant disturbances. This yields
\begin{equation}
	C(z) = \frac{29.504 (z-0.9835)}{z-1} \, ,\quad \hat T_r(z) = \frac{0.025 (z-0.9)}{(z-0.95)^2} \, , \nonumber
\end{equation}
where prefilter 
\begin{equation*}
	V(z) = \frac{0.16527 (z-0.9)}{(z-0.9835)}
\end{equation*}
is used to decrease the overshoot in the step response and therefore to improve the reference tracking performance. The nominal time delay of the plant is $\hat d = 5$; the minimal network delay is assumed to be $\underline\tau^N=0$. The goal is to find the maximal admissible network delay bound $\bar\tau^N$ such that the closed loop is stable. Hence, Theorem~\ref{th:stability_LMI} is evaluated for different $\gamma$ and maximal admissible variable time delays $\bar\tau$ \eqref{eq:tau_bounds}. MOSEK\footnote{https://www.mosek.com (accessed: 19.06.2020)} is used as solver for the involved LMIs.
\begin{figure}[!t]\centering
	\includegraphics[scale=0.4]{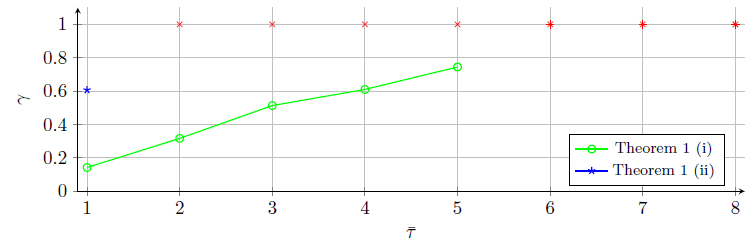}
	\caption{Example, Theorem~\ref{th:stability_LMI}: Comparison of the maximal admissible variable time delay $\bar\tau$ using Theorem~\ref{th:stability_LMI}.}
	\label{fig:LMI_result}
\end{figure}
Figure~\ref{fig:LMI_result} shows the results, where red crosses are used to indicate configurations in which no feasible solution is found. A maximal bound for the time-varying delay $\bar\tau=1$ is found using Theorem~\ref{th:stability_LMI}(ii). The dimension of state vector \eqref{eq:state_vector} is $n_\xi= 9$, which yields a problem with $n_\xi^3 + 2 n_\xi^2 + n_\xi = 900$ optimization variables. In contrast, Theorem~\ref{th:stability_LMI}(i) leads to an optimization problem with $0.5 \big[\big(\hat d +\bar\tau^N\big)^2 + 2 \big(\hat d +\bar\tau^N\big) + 2\big] n_\xi^2 + 0.5 \big( \hat d +\bar\tau^N+2\big) n_\xi = 8046$
variables for a maximal time varying delay of $\bar\tau=8$ as shown in Fig.~\ref{fig:LMI_result}. It is less conservative since it allows to show that a maximal admissible variable time delay is given by $\bar\tau=5$.

\begin{figure}[!t]\centering
	\includegraphics[scale=0.4]{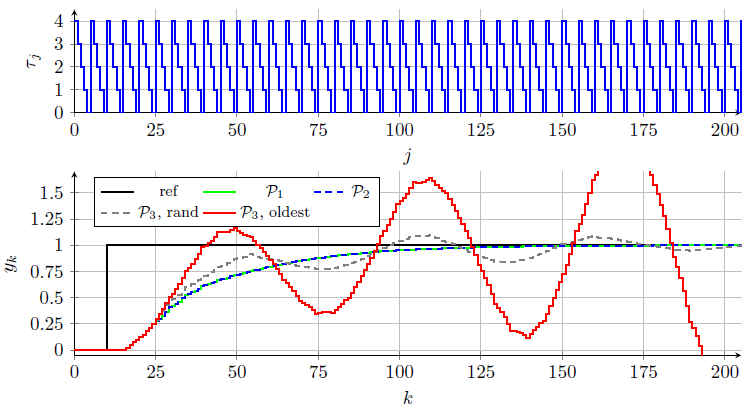}
	\caption{Example, Theorem~\ref{th:stability_LMI}: packet delays $\tau_j$ and output sequences $(y_k)$ for different protocols \eqref{eq:prot}. The red curve represents the worst case scenario.}
	\label{fig:smith_delay_max4}
\end{figure}

A simulation of the closed loop is performed using individual packet delays such that $0 \le \tau_j \le 4 $, i.e. respecting the maximum of $\bar\tau=5$ provided by Theorem~\ref{th:stability_LMI}. Figure~\ref{fig:smith_delay_max4} shows the pattern of the packet delays (top plot), the reference signal (bottom plot, black) and the resulting output sequences, where the three different protocols \eqref{eq:prot} are used in simulation by means of the framework presented in \cite{Steinberger2020}.
It can be clearly seen that the control performance significantly depends on the employed protocol. Especially the use of protocol $\mathcal{P}_3$ yields an unstable loop, if the oldest packet is selected whenever more packets are available at the same time instant. This constitutes the worst case scenario if a packet numbering mechanism is absent as shown in Appendix~\ref{sec:proof}.

This main issue arises because the actual network protocol is not included in the stability analysis, which is, to the best of the authors knowledge, common in LMI-based approaches in literature and is not a specific feature of the presented theorem. Consequently, an alternative way to show the stability of feedback loops with packetized network transmissions is necessary.

\section{Stability Analysis based on the Small Gain Theorem}
\label{sec:SGT}

A way to include protocols like $\mathcal{P}_1$ and $\mathcal{P}_2$ has been proposed in \cite{Steinberger_2020_LCSS}. This work is extended in the following for networked feedback loops with filtered Smith predictors (see Fig.~\ref{fig:filtered_smith_nw}) and protocols $\mathcal{P}_3$, where no numbering of individual packets is implemented.

\subsection{Stability criterion for the case with known plant model}

\begin{figure}[!t]\centering
	\includegraphics[scale=0.4]{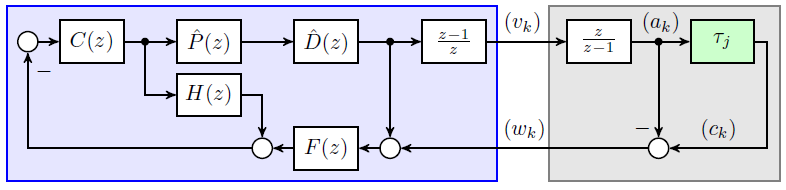}
	\caption{Separation of the feedback loop into a nominal part (blue) and a block that characterizes the uncertainty (gray) due to packet transmissions.}
	\label{fig:smith_SGT}
\end{figure}

To formulate the stability criterion, the networked loop consisting of the filtered Smith predictor, the plant and the transmission channel as in Fig.~\ref{fig:filtered_smith_nw} is rearranged to get Fig.~\ref{fig:smith_SGT}. 
The inputs sequences $(r_k)$ and $(w_k)$ are assumed to be identically to zero for all $k$. The constant time delay \eqref{eq:Dz} is included in the nominal part (blue), the remaining variable time delay \eqref{eq:tau_bounds} is incorporated in the gray part that represents the uncertainty. A discrete-time differentiator and integrator are used to reduce the conservatives in the resulting theorem, as pointed out in \cite{Steinberger_2020_LCSS}. The small gain theorem (SGT) \cite{Sastry1999} forms the basis to deduce the following theorems. Therefore, some basic definitions are recalled. 
\begin{definition}[Truncation of a sequence]
	The truncation of sequence $(f_k)$ at time $T$ is given by $\big(f_k\big)_T = f_k$ for $0 \le k \le T$ and $\big(f_k\big)_T = 0$ for $k > T$.
\end{definition}
\begin{definition}[Extended $\ell_p$ space]
	A sequence $(f_k)$ belongs to the $\ell_p$ space if $\sum_{k=0}^{\infty} \big|f_k\big|^p < \infty$ holds.
	It belongs to the extended $\ell_p$ space, i.e. $(f_k)\in\ell_{pe}$, if $\big(f_k\big)_T\in\ell_{p}$ for all $T$.
\end{definition}
\begin{definition}[Finite gain $\ell_p$ stability]
	\label{def:finite_lp_stability}
	A causal mapping $y_k=\mathcal{M}(u_k): \ell_{pe} \mapsto\ell_{pe}$ is called finite gain $\ell_p$ stable if
	\begin{enumerate}
	\item[(a)] for a given $(u_k)\in\ell_p$, it follows that $(y_k)\in\ell_p$ and
	\item[(b)] for a given $(u_k)\in\ell_{pe}$, it implies that there exist constants $\alpha, \beta >0$ such that there is an affine bound of the norm of output sequence $(y_k)$, i.e.
	$
	\big|\big|\big(y_k\big)\big|\big|_{p,T} \le \alpha 	\big|\big|\big(u_k\big)\big|\big|_{p,T} + \beta 
	$, 
	$\forall T>0$, where $||(\cdot)||_{p,T}$ denotes the p-norm of a truncated sequence $(\cdot)_T$. Constant $\alpha$ is referred to as finite $\ell_p$ gain.
	\end{enumerate}	
\end{definition}
Based on that, we can formulate the following theorem.

\begin{theorem}[Stability criterion, nominal plant]\label{th:stability_SGT}
	Consider the networked filtered Smith predictor as shown in Fig.~\ref{fig:filtered_smith_nw}, where transfer functions $V(z)$ and $C(z)$ are designed to stabilize the nominal, delay-free Plant $\hat P(z)$. The plant \eqref{eq:Pz} is assumed to be known, i.e. $\delta P =0$. Transfer functions $F(z)$ and $H(z)$ are selected according to \eqref{eq:Fz}, and \eqref{eq:Hz}, \eqref{eq:design_Hz}, respectively. The transmission network is characterized by bounded time-varying packet delays as stated in \eqref{eq:tau_bounds}.
	Let 
	\begin{equation}
		\Big|\Big| M(z)\Big|\Big|_{\infty} \alpha = \Bigg|\Bigg| \frac{F(z) C(z) \hat P(z)}{1+C(z) \hat P(z)} \frac{(z-1)}{z}\Bigg|\Bigg|_{\infty} \alpha <1
	\label{eq:th_cond}
	\end{equation}
	hold. The finite $\ell_2$ gain $\alpha$ of the uncertainty is given depending on the network protocol \eqref{eq:prot} so that for
	\begin{enumerate}
		\item[(a)] protocol $\mathcal{P}_1$ (skip old packets, take newest):
		\begin{subequations}\label{eq:th_alphas}
		\begin{equation}
			\alpha = \bar\tau \label{eq:th_alpha_skip} \, ,
		\end{equation}
		\item[(b)] protocol $\mathcal{P}_2$ (do \emph{not} skip old packets, take newest)
		\begin{equation}
			\alpha = \max\Bigg\{ \sqrt{\frac{\bar\tau \left(14\bar\tau^2 -9\bar\tau +1 \right)}{6 (\bar\tau+1)}},1\Bigg\} \label{eq:th_alpha_noskip} \, ,
		\end{equation}
		\item[(c)] protocol $\mathcal{P}_3$ (no numbering nor synchronization)
		\begin{equation}
				\alpha = \sqrt{\frac{\bar\tau}{6} \left(14 \bar\tau + 1\right)} \label{eq:th_alpha_noskip_nonr} \, .
		\end{equation}
		\end{subequations}
	\end{enumerate}	
	Then, the networked feedback loop is finite gain $\ell_2$ stable for all time-varying bounded packet delays $0 \le \underline\tau^N \le \tau_j^N \le \bar\tau^N$.
\end{theorem}

\begin{IEEEproof}
	A proof is given in Appendix~\ref{sec:proof}.
\end{IEEEproof}

\begin{remark}
	Please note that only known information is used in \eqref{eq:th_cond} to obtain the infinity norm. Condition \eqref{eq:th_cond} can be checked easily, e.g., using bode magnitude plots as shown in the example below.
\end{remark}

\begin{remark}
	In \eqref{eq:th_alpha_skip} and \eqref{eq:th_alpha_noskip}, the term ``skip old packets'' means that arriving packets are skipped, whenever more recently sent packets are available at the receiver. Phrase ``take newest'' means that the most recent packet is selected if more packets are received at the controller side at the same time instant. In contrast to cases (a) and (b) in Theorem~\ref{th:stability_SGT}, no numbering or synchronization mechanism is needed for case (c). Gain \eqref{eq:th_alpha_noskip_nonr} is the worst case gain for any packet selection mechanism as can be seen in detail in Appendix~\ref{sec:proof}. 
\end{remark}

\begin{remark}
	Note that the gain for $\mathcal{P}_2$ evaluated at $\bar\tau=1$ equals \eqref{eq:th_alpha_skip} because a packet arrival pattern without overtaking yields the maximal gain in this case. This was not explicitly stated in \cite{Steinberger_2020_LCSS}.
\end{remark}

\subsection{Simulation example: Stability analysis using Theorem~\ref{th:stability_SGT}}

\begin{figure}[!t]
	\includegraphics[scale=0.4]{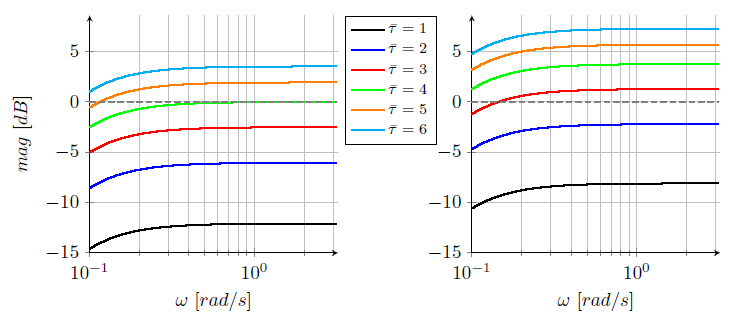}
	\caption{Simulation example, Theorem~\ref{th:stability_SGT}: bode magnitude plots of $M(z) \alpha$ in condition~\eqref{eq:th_cond} for protocol $\mathcal{P}_1$ (left) and $\mathcal{P}_3$ (right).}
	\label{fig:bode_skip_noskip}
\end{figure}

\begin{figure}[!t]
	\includegraphics[scale=0.4]{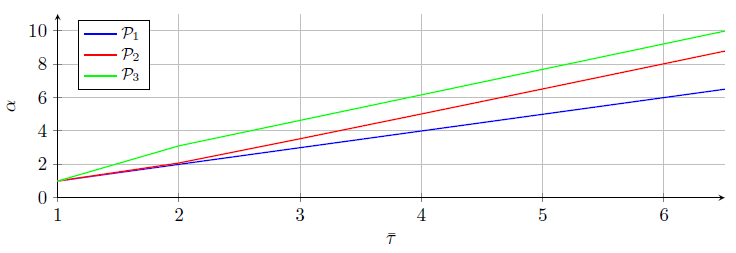}
	\caption{Example, Theorem~\ref{th:stability_SGT}: Comparison of the $\ell_2$ gains as a function of the maximal admissible variable time delay $\bar\tau$ for the three protocols corresponding to Definition~\ref{def:protocols}.}
	\label{fig:comp_gains_tau}
\end{figure}

\begin{figure}[!t]
	\includegraphics[scale=0.4]{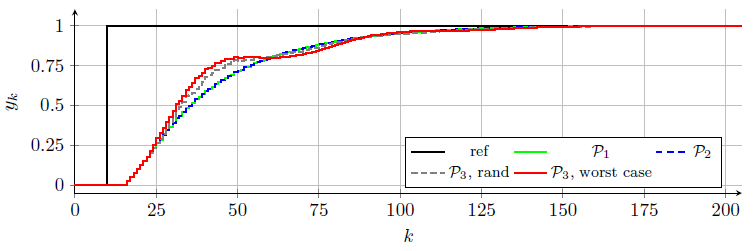}
	\caption{Example, Theorem~\ref{th:stability_SGT}: step responses of the closed networked loop as in Fig.~\ref{fig:filtered_smith_nw} for different protocols \eqref{eq:prot} and a network induced packet pattern as in Fig.~\ref{fig:smith_delay_max4} with $\bar\tau=2$.}
	\label{fig:smith_delay_rand_max2}
\end{figure}

Example 1 from Section~\ref{sec:example1} is revisited in order to show the application of the proposed stability criterion stated in Theorem~\ref{th:stability_SGT}. To do so, condition~\eqref{eq:th_cond} is evaluated for different maximal admissible variable time delays $\bar\tau$ and for protocols $\mathcal{P}_1$ and $\mathcal{P}_3$ exemplarily. Figure~\ref{fig:bode_skip_noskip} presents the corresponding magnitude plots of $M(z) \alpha$ using $\ell_2$ gains \eqref{eq:th_alpha_skip} and \eqref{eq:th_alpha_noskip_nonr}, respectively.
A maximum of $\bar\tau=4$ is achieved for protocol $\mathcal{P}_1$, and $\bar\tau=2$ follows for $\mathcal{P}_3$ because the corresponding magnitudes plot have to stay below $0\, dB$.
The protocol dependent $\ell_2$ gains as stated in \eqref{eq:th_alphas} are depicted in Fig.~\ref{fig:comp_gains_tau}. 
Figure~\ref{fig:smith_delay_rand_max2} shows the achieved step responses for changes in the reference signal $(r_k)$ of the networked filtered Smith predictor in closed loop with plant \eqref{eq:ex_plant} with $\tau_j$ as shown in Fig.~\ref{fig:smith_delay_max4}. Two different scenarios are chosen for $\mathcal{P}_3$ such that the packet leading to the worst case and a random packet are employed, respectively.

\subsection{Stability criterion for the case with uncertain plant model}

\begin{figure}[!t]
	\includegraphics[scale=0.4]{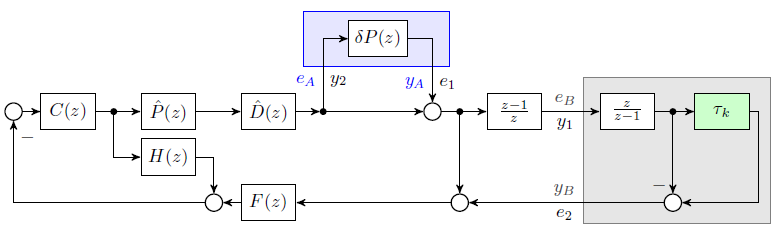}
	\caption{Decomposition of the feedback loop into a nominal part, one uncertainty related to time-varying packet delays (as in Fig.~\ref{fig:smith_SGT}; gray) and a second block $\delta P$ accounting for plant uncertainties (blue).}
	\label{fig:smith_SGT_uncertain}
\end{figure}

\begin{figure}[!t]
	\includegraphics[scale=0.4]{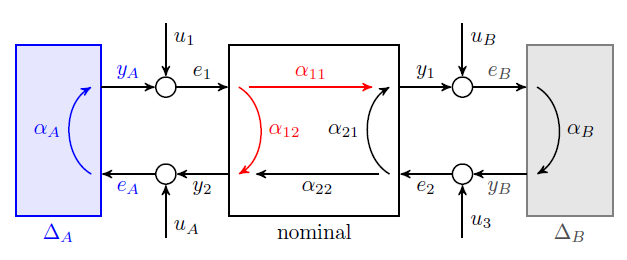}
	\caption{Three block structure consisting of a nominal block, a block
	 $\Delta_B$ including the networked induced variable time delays (gray) and a block $\Delta_A$ representing the plant uncertainties (blue).}
	\label{fig:SGT_3blocks}
\end{figure}

Theorem~\ref{th:stability_SGT} can be extended to also include uncertainties in the plant, i.e. $\delta P \neq 0$. The underlying stability analysis is then based on the block diagram in Fig.~\ref{fig:smith_SGT_uncertain}, which is a modification of the structure shown in Fig.~\ref{fig:smith_SGT}.
Using sequences $(e_A)$, $(y_A)$ as well as $(e_B)$, $(y_B)$, Fig.~\ref{fig:smith_SGT_uncertain} can be drawn as depicted in Fig.~\ref{fig:SGT_3blocks}. Here, additional input signals $(u_1)$, $(u_3)$, $(u_A)$ and $(u_B)$ are introduced and exploited in the proof of the following theorem. 
The $\alpha$'s stand for the various $\ell_2$ gains between different inputs and outputs of the three considered blocks.

\begin{theorem}[Stability criterion, uncertain plant]\label{th:stability_SGT_uncertain}
	Let the same assumptions hold as in Theorem~\ref{th:stability_SGT} except that plant model \eqref{eq:Pz} is uncertain such that $\big|\big|\delta P\big|\big|_\infty=\alpha_A > 0$. In addition, conditions
	\begin{subequations}\label{eq:th_uncertain_conditions}
	\begin{align}
		\alpha_A \alpha_{12} + \frac{\alpha_A \alpha_B \alpha_{11} \alpha_{22}}{1-\alpha_B \alpha_{21}} &< 1 \ , \quad \alpha_B \alpha_{21} < 1 \label{eq:th_uncertain_cond1}\\
		\alpha_B \alpha_{21} + \frac{\alpha_A \alpha_B \alpha_{11} \alpha_{22}}{1-\alpha_A \alpha_{12}} &< 1 \ , \quad \alpha_A \alpha_{12} < 1 
	\end{align}
	\end{subequations}
	hold, where
	\begin{subequations}\label{eq:th_uncertain_alphas}
	\begin{align}
		\alpha_{11} &= \bigg|\bigg| \hat S(z) \frac{(z-1)}{z}\bigg|\bigg|_{\infty}, \qquad\;
		\alpha_{12} = \Big|\Big| F(z) \hat T(z)\Big|\Big|_{\infty}\\
		\alpha_{21} &= \bigg|\bigg| F(z) \hat T(z) \frac{(z-1)}{z}\bigg|\bigg|_{\infty}, \, 
		\alpha_{22} = \Big|\Big| F(z) \hat T(z) \Big|\Big|_{\infty}\\
		\hat T(z) &= \frac{C(z) \hat P(z)}{1+C(z) \hat P(z)} , \quad\ \hat S(z) = \frac{1}{1+C(z) \hat P(z)}
	\end{align}
	\end{subequations}
	and $\alpha_A$ equals \eqref{eq:th_alphas}, depending on the chosen protocol~\eqref{eq:prot}.\\
	Then, the feedback loop consisting of the filtered Smith predictor, the uncertain plant and the packetized transmission network is finite gain $\ell_2$ stable for all time-varying bounded packet delays $0 \le \underline\tau^N \le \tau_j^N \le \bar\tau^N$.
\end{theorem}
\begin{IEEEproof}
	A proof is given in Appendix~\ref{sec:proof_uncertain}.
\end{IEEEproof}
\begin{remark}
	Please note that conditions \eqref{eq:th_uncertain_conditions}, \eqref{eq:th_uncertain_alphas} collapse into~\eqref{eq:th_cond} of Theorem~\ref{th:stability_SGT} for the case without uncertainties in the plant, i.e. $\delta P=0$, or equivalently, $\alpha_A=0$. Similar to this, conditions \eqref{eq:th_uncertain_conditions}, \eqref{eq:th_uncertain_alphas} reduce to the classical formulation of the SGT \cite{Sastry1999} applied to  the filtered Smith predictor if no time-varying delays are present. In this case, the green block in Fig.~\ref{fig:smith_SGT_uncertain} representing the variable time delay is replaced by a direct connection resulting in $\alpha_B=0$.
\end{remark}

\section{Summary and Outlook}
\label{sec:conclusion}

Criteria for the stability analysis of networked Smith predictors are proposed. The main ingredient is the calculation of the $\ell_2$ gain of the uncertainty caused by the bounded but time-varying network delays, which depends on the actual protocol of the packetized transmission network. It is pointed out that the explicit consideration of the effects of packet skipping and hold mechanisms have to be explicitly taken into account. However, this is commonly neglected in LMI-based approaches as was shown using a simple simulation example.
One additional benefit of the introduced theorems is that they can be checked easily, e.g., using bode magnitude plots without involving any optimization problem.

Future extensions may deal with the inclusion of additional protocols and mechanisms to deal with packet dropouts. Also a generalization to multivariable systems might be of interest.

\appendices
\section{Proof of Theorem~\ref{th:stability_SGT}}
\label{sec:proof}

\begin{figure}[!t]\centering
	\includegraphics[scale=0.4]{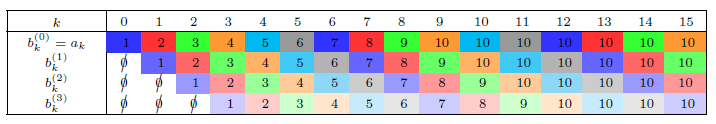}
	\caption{Representation of the packets ``on the way'' for the case with $0 \le \tau_j \le 3$ and $T=9$. Lighter colors represent older packets.}
	\label{fig:packets_general_x10}
\end{figure}

\begin{figure}[!t]\centering
	\includegraphics[scale=0.4]{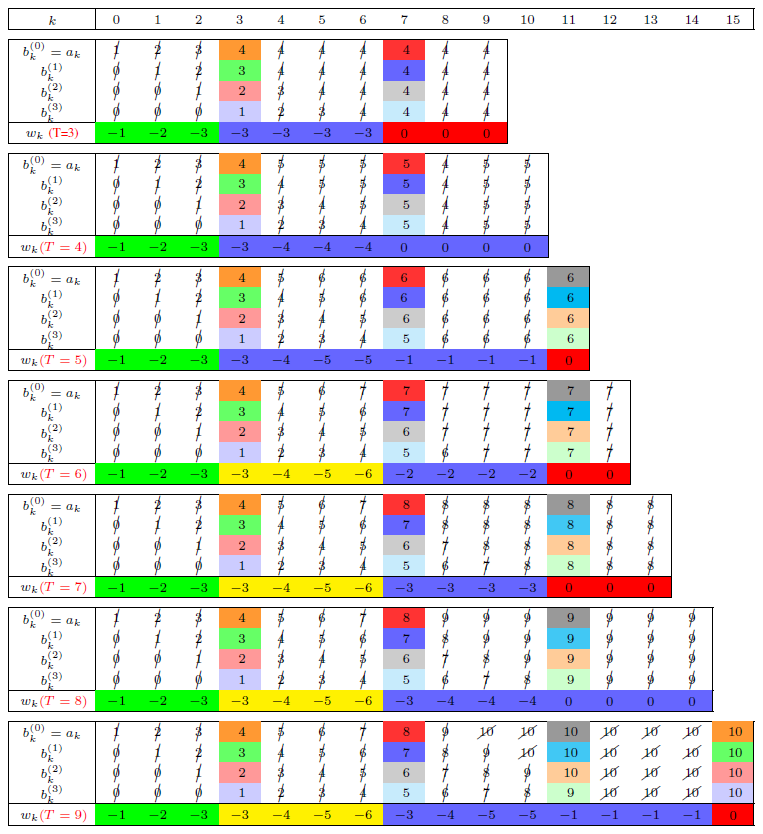}
	\caption{Worst case patterns of arriving packets for the case with $0 \le \tau_j \le \bar\tau=3$ and $T\in\{3,4,5,6,7,8,9\}$.}
	\label{fig:packets_noskip_nonr_all}
\end{figure}

\begin{figure}[!t]\centering
	\includegraphics[scale=0.4]{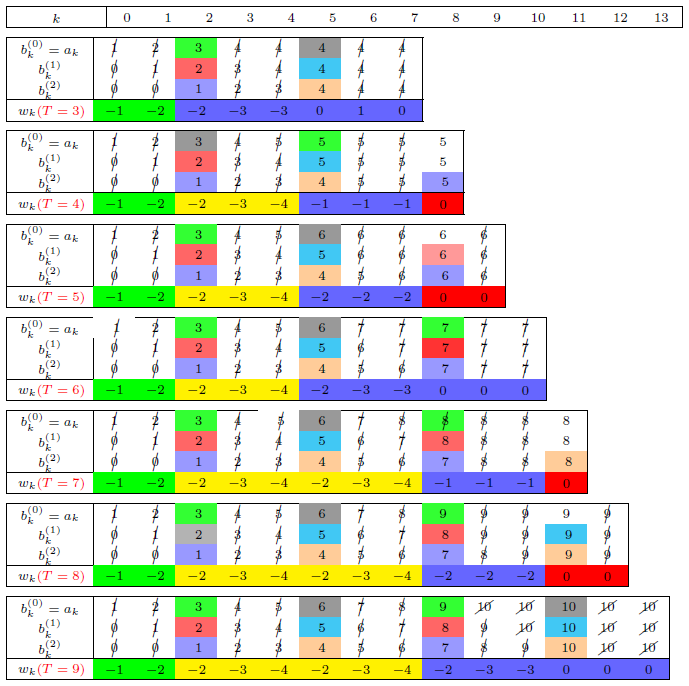}
	\caption{Worst case patterns of arriving packets for the case with $0 \le \tau_j \le \bar\tau=2$ and $T\in\{3,4,5,6,7,8,9\}$.}
	\label{fig:packets_noskip_nonr_tau2_all}
\end{figure}

The nominal (blue) part in Fig.~\ref{fig:smith_SGT} with input $(w_k)$ and output $(v_k)$ can be described by transfer function
\begin{equation}
	\hat T_{v}(z) = \frac{\mathcal{Z}\big\{(v_k)\big\}}{\mathcal{Z}\big\{(w_k)\big\}}  = -\frac{z-1}{z} \frac{F(z) C(z) \hat P(z) \hat D(z)}{1+C(z) \hat P(z)} \, .
\end{equation}
Applying the SGT \cite{Sastry1999} to the structure in Fig.~\ref{fig:smith_SGT} yields condition \eqref{eq:th_cond}, where the fact that $||D(z)||_\infty=1$ was exploited. 
The $\ell_2$ gains $\alpha$ of the uncertainty depend on the actually used protocols, see Definition~\ref{def:protocols}. As shown in \cite{Steinberger_2020_LCSS}, the gains for protocol $\mathcal{P}_1$ and $\mathcal{P}_2$ are given by \eqref{eq:th_alpha_skip} and \eqref{eq:th_alpha_noskip}, respectively. Hence, only \eqref{eq:th_alpha_noskip_nonr}, i.e. the $\ell_2$ gain of the gray block in Fig.~\ref{fig:smith_SGT} for protocol $\mathcal{P}_3$, has to be shown next. Thus, the 2-norm of output $w_k = c_k - a_k$ has to be maximized. Following the idea from \cite{Steinberger_2020_LCSS}, a truncated input sequence $(v_k) = \big(v_0, v_1, v_2, \ldots, v_T, 0, \ldots\big) = \big(\bar v, \bar v, \bar v, \ldots, \bar v, 0,  \ldots\big)$ is defined for time instants $k=(0, 1, 2, \ldots)$. Because of the discrete-time integrator in Fig.~\ref{fig:smith_SGT}, one gets $(a_k) = \big(a_0, a_1, a_2, \ldots, a_T, a_{T+1}, \ldots\big) 	= \big(\bar v, 2\bar v, 3\bar v, \ldots, (T+1)\bar v, (T+1)\bar v, \ldots\big)$ that is directly fed to output $w_k$. In addition, sequence $(a_k)$ passes the green block in Fig.~\ref{fig:smith_SGT}, which represents the packetized transmissions that are subject to time-varying delays. It comprises a transmitter as well as a packet selection and a hold mechanism as specified in \eqref{eq:prot}. Each packet $a_k$ might be delayed between $0$ and $\bar\tau$ time steps as indicated by $b_k^{(0)}, b_k^{(1)}, \ldots, b_k^{(\bar\tau)}$, respectively. This situation with several packets ``on the way'' is exemplified in Fig.~\ref{fig:packets_general_x10} for $\bar\tau=3$ and $T=9$.
A packet containing $a_k$ arrives at the latest after $\bar\tau$ time instants. Depending on the actual packet selecting mechanism in \eqref{eq:p_worst}, the sample $c_k$ corresponding to the chosen packet is provided at the output of the (green) delay block in Fig.~\ref{fig:smith_SGT}. A hold mechanism keeps the previously selected sample, if no new packet arrives, see the definition of $\mathcal{P}_3$ in \eqref{eq:prot}. To show that the worst case $\ell_p$ gain of the uncertainty for $\mathcal{P}_3$ is \eqref{eq:th_alpha_noskip_nonr}, one has to maximize the squared 2-norm of output $w_k$. Thus,  $T+1+2\bar\tau$ samples have to be considered because they contribute to $w_k=a_k-c_k$. Values $a_k$ with  $k \ge T+1+2\bar\tau$ do not affect $w_k$, cf. Fig.~\ref{fig:packets_noskip_nonr_all} and \ref{fig:packets_noskip_nonr_tau2_all}. All initial conditions, as e.g. $c_k$ for $k=-1$, are assumed to be zero.

\begin{figure}[!t]\centering
	\includegraphics[scale=0.4]{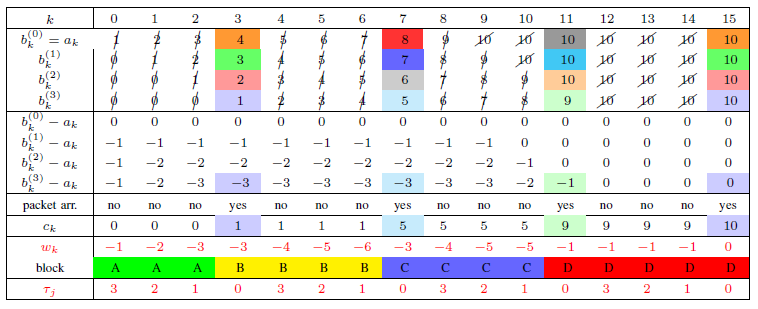}
	\caption{Pattern of arriving packets (represented using different colors) that lead to the worst case gain for the case with $0 \le \tau_j \le 3$ and $T=9$.}
	\label{fig:packets_noskip_nonr_x10}
\end{figure}

Due to the fact that $(a_k)$ is monotonically increasing and the difference between $a_k$ and $c_k$ is used to get $w_k$, the worst case sequence $(w_k)$ results if one selects the packets containing the smallest $a_k$ and keep them as long as possible. This can be clearly seen using, for example, $\bar\tau=3$ and $T=9$ as in Fig.~\ref{fig:packets_noskip_nonr_x10}.
Sample $a_0=1$ is received at the latest possible time instant $k=\bar\tau=3$, which yield $c_k=b_k^{(3)}-a_k=1$ for $k=3$. No packet has to arrive for $k=0,1,2$ to maximize $(w_k)$. In addition, $c_3=1$ has to be hold as long as possible, i.e. until the next packet has to arrive due to Assumption~\ref{as:delay}. This means, that no packet should arrive at time instants $k=4,5,6$ and $a_4=5$ is selected at $k=7$. Consequently, the packets containing $a_1=2$, $a_2=3$ and $a_3=4$ have to be discarded, which is only possible if they arrive at $k=3$ and are not selected at this time instant, see Fig.~\ref{fig:packets_noskip_nonr_x10}. Next, $c_7=5$ has to be kept as long as possible until packet $a_8=9$ is selected at $k=11$. The entire pattern of received packets, all packets ``on the way'', the resulting worst case sequence $(w_k)$ and the corresponding packet delays $\tau_j$ are depicted in Fig.~\ref{fig:packets_noskip_nonr_x10}.

\begin{figure}[!t]
	\includegraphics[scale=0.4]{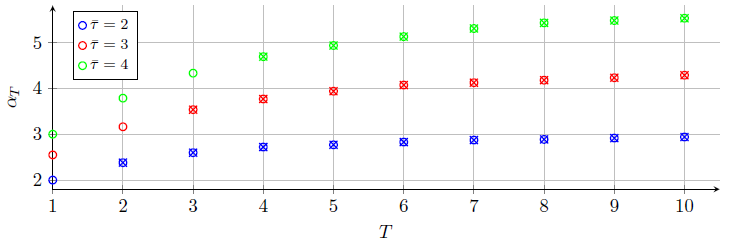}
	\caption{Comparison of actual (circles) and calculated (crosses) gains $\alpha_T$ for different maximal delays $\bar\tau$ and $T+1$ samples of $(v_k)$ different from zero.}
	\label{fig:noskip_nonr_gain}
\end{figure}

\begin{figure}[!t]
	\includegraphics[scale=0.4]{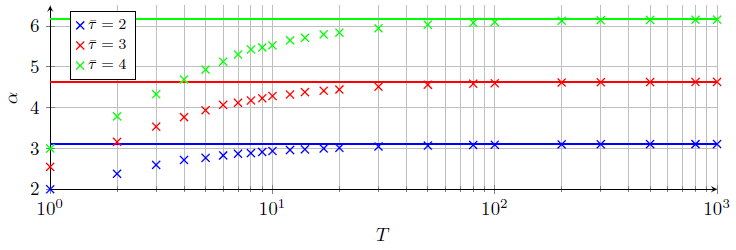}
	\caption{Gains $\alpha_T$ (crosses) and the associated maximal gains  $\alpha$ (solid lines) for different $\bar\tau$.}
	\label{fig:noskip_nonr_gain_long}
\end{figure}

Figures~\ref{fig:packets_noskip_nonr_all} and \ref{fig:packets_noskip_nonr_tau2_all} present the worst case patterns of arriving packets for the cases with different $\bar\tau$ and $T$. The worst case sequence $(w_k)$ consists of maximal four different blocks as indicated in the figures using different colors, see also block A, B, C and D in Fig.~\ref{fig:packets_noskip_nonr_x10}. 
The worst case pattern consists of a (green) introductory part A consisting of $\bar\tau$ samples, followed by the remaining $T+1+\bar\tau$ samples that are split into several parts. First, the $k_1$ (yellow) blocks B with length $\bar\tau+1$ are separated from the rest to get $\le3\bar\tau$ remaining samples. This rest with $k_2$ elements is split into a (blue) block C with length $k_3=\big\lfloor\frac{k_2}{\bar\tau+1}\big\rfloor$, which represents the smallest integer larger than $\frac{k_2}{\bar\tau+1}$, and a (red) block D with $k_2-k_3(\bar\tau+1)$ elements. Note that the last element is always equal to zero. 
Based on this splitting into four parts, on can formalize the 2-norm of the truncated worst case sequence such that
\begin{align}
	\big|\big|(w_k)\big|\big|_{2,T}^2 &= \underbrace{\sum_{i=1}^{\bar\tau} i^2 \bar v^2}_{A} + \label{eq:wk_norm_noskip_nonr}\\
	&\ \underbrace{\sum_{j=0}^{k_1-1} \Bigg\{\sum_{i=\bar\tau+j(\bar\tau+1)}^{2\bar\tau+j(\bar\tau+1)} \left(a_i - a_{j(\bar\tau+1)}\right)^2 \Bigg\}}_{B} + \nonumber\\
	&\ \underbrace{\sum_{j=k_1}^{k_1+k_3-1} \Bigg\{\sum_{i=\bar\tau+j(\bar\tau+1)}^{2\bar\tau+j(\bar\tau+1)} \left(a_i - a_{j(\bar\tau+1)}\right)^2 \Bigg\}}_{C} + \nonumber\\
	& \ \underbrace{\sum_{i=\bar\tau+(k_1+k_3)(\bar\tau+1)}^{T+2\bar\tau} \left(a_i -a_{\bar\tau+(k_1+k_3)(\bar\tau+1)}\right)^2}_{D} \nonumber
\end{align}
with the underlying worst case packet delay pattern
\begin{equation}
	(\tau_j) = \big(\underbrace{\bar\tau, \bar\tau-1, \bar\tau-2, \ldots, 0,}_{\bar\tau+1 \text{ elements}} \underbrace{\bar\tau, \bar\tau-1, \bar\tau-2, \ldots, 0,}_{\bar\tau+1 \text{ elements}}\ \ldots \big) \, .
\end{equation}
Next, the finite $\ell_2$ gain $\alpha$ is calculated such that 
\begin{equation}
	\alpha = \max_{T>0} \alpha_T \qquad\text{with}\qquad 
	\alpha_T = \sqrt{\frac{\big|\big|(w_k)\big|\big|_{2,T}^2}{\big|\big|(v_k)\big|\big|_{2,T}^2}}
	\label{eq:alpha_calc} \, ,
\end{equation}
and $||(v_k)||_{2,T}^2 = (1+T) \bar v^2$, see also Definition~\ref{def:finite_lp_stability}. Figure~\ref{fig:noskip_nonr_gain} compares calculated values for $\alpha_T$ using relations \eqref{eq:alpha_calc}, \eqref{eq:wk_norm_noskip_nonr} with the actual (true) results, which are obtained by evaluating the effect of all possible combinations of bounded packet delays $\tau_j$ on the norm of sequence $(w_k)_T$.
To get the finite $\ell_2$ gain $\alpha$ \eqref{eq:alpha_calc}, the limit of $\alpha_T$ for $T\rightarrow\infty$ is computed using $\big|\big|(w_k)\big|\big|_{2,T}^2 = A + B + C + D$. 
The part corresponding to B consists of $k_1$ blocks of the form $\left(-\bar\tau, -(\bar\tau+1)\bar v, -(\bar\tau+2)\bar v \ldots, -2\bar\tau\bar v \right)$ with $\bar\tau+1$ samples each. One of these blocks B contributes to the overall 2-norm of $(w_k)_T$ with
\begin{eqnarray}
	d &=& \sum_{i=1}^{\bar\tau+1} \big((\bar\tau -1) + i\big) \bar v^2 \nonumber\\
	&=& \Bigg( (\bar\tau -1)^2 \sum_{i=1}^{\bar\tau+1} 1 + 2 (\bar\tau -1) \sum_{i=1}^{\bar\tau+1} i + \sum_{i=1}^{\bar\tau+1} i^2 \Bigg)\bar v^2 \nonumber\\
	&=& \frac{\bar\tau+1}{6} \left(14 \bar\tau^2 +\bar\tau\right) \bar v^2 \, .
\end{eqnarray}
leading to
\begin{equation}
	\big|\big|(w_k)\big|\big|_{2,T}^2 = A + \frac{T+1}{\bar\tau+1} d + C + D\, ,
\end{equation}
where blocks A, C and D are constant with respect to $T$. Consequently, one gets
\begin{align}
	\alpha_T^2 &= \frac{A + \frac{T+1}{\bar\tau+1} d + C + D}{1+T} = \frac{A+C+D}{1+T} + \frac{d}{1+T} \ , \\
	\alpha &= \lim_{T\rightarrow\infty} \alpha_T = \sqrt{\frac{d}{\bar\tau+1}} = \sqrt{\frac{14 \bar\tau^2}{6} +\frac{\bar\tau}{6}} \, ,
\end{align} 
see \eqref{eq:th_alpha_noskip_nonr} in Theorem~\ref{th:stability_SGT}. This completes the proof. Figure~\ref{fig:noskip_nonr_gain_long}
exemplifies gains $\alpha_T$ and $\alpha$ for different maximal time-varying delays $\bar\tau$.  \hfill $\blacksquare$

\section{Proof of Theorem~\ref{th:stability_SGT_uncertain}}
\label{sec:proof_uncertain}

The proof follows the same main idea that is usually used to prove the SGT, see, e.g., \cite{Sastry1999}. Hence, the structure in Fig.~\ref{fig:SGT_3blocks} is analyzed using the corresponding $\ell_2$ gains. It is shown that all signals in this three-block structure are bounded as long as all input signals are bounded. For simplicity, the iteration index $k$ is skipped in the proof so that, e.g., the 2-norm of truncated sequence $(y_{A,k})_T$ is written as $\big|\big| y_A \big|\big|_T$. The plant uncertainty and the uncertainty due to the time-varying delay can be cast into a form as in Definition~\ref{def:finite_lp_stability} such that
\begin{equation}
\big|\big| y_A \big|\big|_T \le \alpha_A \big|\big| e_A \big|\big|_T + \beta_A\, , \
\big|\big| y_B \big|\big|_T \le \alpha_B \big|\big| e_B \big|\big|_T + \beta_B 
\end{equation}
hold. Due to linearity, the input-output behavior of the nominal block in Fig.~\ref{fig:SGT_3blocks} can be written as
\begin{subequations}
\begin{align}
	\big|\big| y_1 \big|\big|_T &\le \alpha_{11} \big|\big| e_1 \big|\big|_T+ \beta_{11} + \alpha_{21} \big|\big| e_2 \big|\big|_T + \beta_{21} \, ,\\
	\big|\big| y_2 \big|\big|_T &\le \alpha_{12} \big|\big| e_1 \big|\big|_T + \beta_{12} + \alpha_{22} \big|\big| e_2 \big|\big|_T + \beta_{22}  \, .
\end{align}
\end{subequations}
To prove Theorem~\ref{th:stability_SGT_uncertain}, one has to show that all inner signals, as e.g. $e_1=u_1+y_A$, are bounded. Thus,
\begin{equation}
	\big|\big| e_1 \big|\big|_T \le \big|\big| u_1 \big|\big|_T + \alpha_A \big|\big| e_A \big|\big|_T + \beta_A
\end{equation}
with $e_A=u_A+y_1$ and
$\big|\big| e_A \big|\big|_T \le \big|\big| u_A \big|\big|_T + \alpha_{12} \big|\big| e_1 \big|\big|_T + \beta_{12} + \alpha_{22} \big|\big| e_2 \big|\big|_T + \beta_{22}
$ is used. Incorporating $e_2=u_2+y_B$ and $e_B=u_B+y_1$ yields
\begin{equation}
	\big|\big| e_2 \big|\big|_T \left( 1-\alpha_B \alpha_{21}\right) \le \delta_1 + \alpha_B \alpha_{11} \big|\big| e_1 \big|\big|_T
\end{equation}
with
$\delta_1 = \big|\big| u_2 \big|\big|_T + \alpha_B \big|\big| u_B \big|\big|_T  + \beta_B +\alpha_B \left(\beta_{11}+\beta_{21}\right)$ and so
\begin{equation}
	\big|\big| e_1 \big|\big|_T \left( 1-\alpha_A \alpha_{21} - \frac{\alpha_A \alpha_B \alpha_{11} \alpha_{22}}{1-\alpha_B \alpha_{21}}\right) \le \delta_2 
\end{equation}
where
$\delta_2 = \big|\big| u_1 \big|\big|_T + \beta_A + 
	\alpha_A \big(\big|\big|u_A \big|\big|_T + \beta_{12}+\beta_{22} + \frac{\alpha_{22}\delta_1}{1-\alpha_B \alpha_{21}}\big)$.
As a result, error $e_1$ is norm-bounded for bounded input signals $||u_1||_T$, $||u_2||_T$, $||u_A||_T$ and $||u_B||_T$ if
\begin{equation}
	1-\alpha_A \alpha_{21} - \frac{\alpha_A \alpha_B \alpha_{11} \alpha_{22}}{1-\alpha_B \alpha_{21}} >0 \quad \text{for}\quad 
	1-\alpha_B \alpha_{21} >0 
\end{equation}
that constitute conditions~\eqref{eq:th_uncertain_cond1} in Theorem~\ref{th:stability_SGT_uncertain}. Analogue steps are followed to show that the remaining $|| e_i||_T$ and $|| y_i||_T$, $i\in\{1,2,A,B\}$, are bounded for bounded inputs $|| u_i||_T$.
The gains $\alpha_{11}$, $\alpha_{12}$, $\alpha_{21}$ and $\alpha_{22}$ \eqref{eq:th_uncertain_alphas} directly follow by computing down the transfer functions corresponding to Fig.~\ref{fig:smith_SGT_uncertain}.
This completes the proof. \hfill $\blacksquare$

\bibliographystyle{IEEEtran}
\bibliography{Smith_var_delay_arxiv}

\begin{thebibliography}{10}
\providecommand{\url}[1]{#1}
\csname url@samestyle\endcsname
\providecommand{\newblock}{\relax}
\providecommand{\bibinfo}[2]{#2}
\providecommand{\BIBentrySTDinterwordspacing}{\spaceskip=0pt\relax}
\providecommand{\BIBentryALTinterwordstretchfactor}{4}
\providecommand{\BIBentryALTinterwordspacing}{\spaceskip=\fontdimen2\font plus
\BIBentryALTinterwordstretchfactor\fontdimen3\font minus
  \fontdimen4\font\relax}
\providecommand{\BIBforeignlanguage}[2]{{%
\expandafter\ifx\csname l@#1\endcsname\relax
\typeout{** WARNING: IEEEtran.bst: No hyphenation pattern has been}%
\typeout{** loaded for the language `#1'. Using the pattern for}%
\typeout{** the default language instead.}%
\else
\language=\csname l@#1\endcsname
\fi
#2}}
\providecommand{\BIBdecl}{\relax}
\BIBdecl

\bibitem{Gupta2010}
R.~A. Gupta and M.-Y. Chow, ``{Networked Control System: Overview and Research
  Trends},'' \emph{{IEEE Trancactions on Industial Electronics}}, vol.~{57},
  no.~{7}, pp. {2527--2535}, {2010}.

\bibitem{Hespanha2007}
J.~P. Hespanha, P.~Naghshtabrizi, and Y.~Xu, ``{A Survey of Recent Results in
  Networked Control Systems},'' \emph{{Proceedings of the IEEE}}, vol.~{95},
  no.~{1}, pp. {138--162}, {2007}.

\bibitem{Zhang2017}
D.~Zhang, P.~Shi, Q.-G. Wang, and L.~Yu, ``{Analysis and synthesis of networked
  control systems: A survey of recent advances and challenges},'' \emph{{ISA
  Transactions}}, vol.~{66}, pp. {376--392}, {2017}.

\bibitem{Park2018}
P.~Park, S.~C. Ergen, C.~Fischione, C.~Lu, and K.~H. Johansson, ``{Wireless
  Network Design for Control Systems: A Survey},'' \emph{{IEEE Communications
  Surveys and Tutorials}}, vol.~{20}, no.~{2}, pp. {978--1013}, {2018}.

\bibitem{Schenato2007}
L.~{Schenato}, B.~{Sinopoli}, M.~{Franceschetti}, K.~{Poolla}, and S.~S.
  {Sastry}, ``Foundations of control and estimation over lossy networks,''
  \emph{Proceedings of the IEEE}, vol.~95, no.~1, pp. 163--187, 2007.

\bibitem{Palmisano2021}
M.~Palmisano, M.~Steinberger, and M.~Horn, ``Optimal finite-horizon control for
  networked control systems in the presence of random delays and packet
  losses,'' \emph{IEEE Control Systems Letters}, vol.~5, no.~1, pp. 271 -- 276,
  2021.

\bibitem{Heemels2010}
W.~P. M.~H. Heemels and N.~van~de Wouw, ``Stability and stabilization of
  networked control systems,'' in \emph{Networked Control Systems}, ser.
  Lecture Notes in Control and Information Sciences, {Bemporad A., Heemels M.,
  Johansson M.}, Ed.\hskip 1em plus 0.5em minus 0.4em\relax Springer, London,
  2010, vol. 406, pp. 203--253.

\bibitem{Liu2010}
G.~P. Liu, ``Predictive controller design of networked systems with
  communication delays and data loss,'' \emph{{IEEE} Transactions on Circuits
  and Systems {II}: Express Briefs}, vol.~57, no.~6, pp. 481--485, 2010.

\bibitem{Mu2005}
J.~Mu, G.~Liu, and D.~Rees, ``{Design of robust networked predictive control
  systems},'' in \emph{{ACC: Proceedings of the 2005 American Control
  Conference, Vols 1-7}}, {2005}, pp. {638--643}.

\bibitem{Cloosterman2010}
M.~Cloosterman, L.~Hetel, N.~van~de Wouw, W.~Heemels, J.~Daafouz, and
  H.~Nijmeijer, ``Controller synthesis for networked control systems,''
  \emph{Automatica}, vol.~46, no.~10, pp. 1584 -- 1594, 2010.

\bibitem{Li2011b}
H.~Li, Z.~Sun, M.-Y. Chow, and F.~Sun, ``{Gain-Scheduling-Based State Feedback
  Integral Control for Networked Control Systems},'' \emph{{IEEE Transactions
  on Industrial Electronics}}, vol.~{58}, no.~{6}, pp. {2465--2472}, {2011}.

\bibitem{Ludwiger2018}
J.~Ludwiger, M.~Steinberger, M.~Horn, G.~Kubin, and A.~Ferrara, ``Discrete time
  sliding mode control strategies for buffered networked systems,'' in
  \emph{57th Annual Conference on Decision and Control (CDC)}, 2018, pp.
  6735--6740.

\bibitem{Ludwiger2019}
J.~{Ludwiger}, M.~{Steinberger}, and M.~{Horn}, ``Spatially distributed
  networked sliding mode control,'' \emph{IEEE Control Systems Letters},
  vol.~3, no.~4, pp. 972--977, 2019.

\bibitem{Incremona2017}
G.~P. Incremona, A.~Ferrara, and L.~Magni, ``Asynchronous networked {MPC} with
  {ISM} for uncertain nonlinear systems,'' \emph{IEEE Transactions on Automatic
  Control}, vol.~62, no.~9, pp. 4305--4317, 2017.

\bibitem{Smith1959}
O.~Smith, ``Closer control of loops with dead time,'' \emph{Chemical
  Engineering Progress}, vol.~53, no.~5, pp. 217--219, 1959.

\bibitem{Normey2009}
J.~E. Normey-Rico and E.~F. Camacho, ``Unified approach for robust dead-time
  compensator design,'' \emph{Journal of Process Control}, vol.~19, no.~1, pp.
  38 -- 47, 2009.

\bibitem{Santos2016}
T.~L.~M. Santos, B.~C. Torrico, and J.~E. Normey-Rico, ``{Simplified filtered
  Smith predictor for MIMO processes with multiple time delays},'' \emph{{ISA
  Transactions}}, vol.~{65}, pp. {339--349}, {2016}.

\bibitem{Lai2010}
C.-L. Lai and P.-L. Hsu, ``{Design the Remote Control System With the
  Time-Delay Estimator and the Adaptive Smith Predictor},'' \emph{{IEEE
  Transactions on Industrial Electronics}}, vol.~{6}, no.~{1}, pp. {73--80},
  {2010}.

\bibitem{Repele2014}
L.~Repele, R.~Muradore, D.~Quaglia, and P.~Fiorini, ``{Improving Performance of
  Networked Control Systems by Using Adaptive Buffering},'' \emph{{IEEE
  Transactions on Industrial Electronics}}, vol.~{61}, no.~{9}, pp.
  {4847--4856}, {2014}.

\bibitem{Gamal2016}
M.~Gamal, N.~Sadek, M.~R.~M. Rizk, and A.~K. Abou-elSaoud, ``{Delay
  compensation using Smith predictor for wireless network control system},''
  \emph{{Alexandria Engineering Journal}}, vol.~{55}, no.~{2}, pp.
  {1421--1428}, {2016}.

\bibitem{Batista2018}
A.~P. Batista and P.~G. Jota, ``{Performance improvement of an NCS closed over
  the internet with an adaptive Smith Predictor},'' \emph{{Control Engineering
  Practice}}, vol.~{71}, pp. {34--43}, {2018}.

\bibitem{Bonala2017}
S.~Bonala, B.~Subudhi, and S.~Ghosh, ``{On delay robustness improvement using
  digital Smith predictor for networked control systems},'' \emph{{European
  Journal of Control}}, vol.~{34}, pp. {59--65}, {2017}.

\bibitem{Normey2012}
J.~E. Norrney-Rico, P.~Garcia, and A.~Gonzalez, ``{Robust stability analysis of
  filtered Smith predictor for time-varying delay processes},'' \emph{Journal
  of Process Control}, vol.~{22}, no.~{10}, pp. {1975--1984}, {2012}.

\bibitem{Steinberger2020}
M.~Steinberger, M.~Tranninger, M.~Horn, and K.~H. Johansson, ``How to simulate
  networked control systems with variable time delays?'' in \emph{21st IFAC
  World Congress}, Berlin, 2020.

\bibitem{Mathworks2012}
F.~Zhang and M.~Yeddanapudi, ``Modeling and simulation of time-varying
  delays,'' in \emph{Proceedings of the 2012 Symposium on Theory of Modeling
  and Simulation}, 2012.

\bibitem{Cervin2003}
A.~{Cervin}, D.~{Henriksson}, B.~{Lincoln}, J.~{Eker}, and K.~. {Arzen}, ``{How
  does control timing affect performance? Analysis and simulation of timing
  using Jitterbug and TrueTime},'' \emph{IEEE Control Systems Magazine},
  vol.~23, no.~3, pp. 16--30, 2003.

\bibitem{Li2011}
X.~{Li} and H.~{Gao}, ``A new model transformation of discrete-time systems
  with time-varying delay and its application to stability analysis,''
  \emph{IEEE Transactions on Automatic Control}, vol.~56, no.~9, pp.
  2172--2178, 2011.

\bibitem{Seuret2015}
A.~{Seuret}, F.~{Gouaisbaut}, and E.~{Fridman}, ``Stability of discrete-time
  systems with time-varying delays via a novel summation inequality,''
  \emph{IEEE Transactions on Automatic Control}, vol.~60, no.~10, pp.
  2740--2745, 2015.

\bibitem{Liu2015}
A.~Liu, W.-a. Zhang, L.~Yu, S.~Liu, and M.~Z.~Q. Chen, ``{New results on
  stabilization of networked control systems with packet disordering},''
  \emph{{Automatica}}, vol.~{52}, pp. {255--259}, {2015}.

\bibitem{Wu2018}
W.~Wu and Y.~Zhang, ``{Event-triggered fault-tolerant control and scheduling
  codesign for nonlinear networked control systems with medium-access
  constraint and packet disordering},'' \emph{{International Journal of Robust
  and Nonlinear Control}}, vol.~{28}, no.~{4}, pp. {1182--1198}, {2018}.

\bibitem{Kao2004}
C.-Y. Kao and B.~Lincoln, ``Simple stability criteria for systems with
  time-varying delays,'' \emph{Automatica}, vol.~40, no.~8, pp. 1429 -- 1434,
  2004.

\bibitem{Sastry1999}
S.~Sastry, \emph{Nonlinear Systems: Analysis, Stability and Control}.\hskip 1em
  plus 0.5em minus 0.4em\relax Springer, 1999.

\bibitem{Steinberger_2020_LCSS}
M.~Steinberger and M.~Horn, ``A stability criterion for networked control
  systems with packetized transmissions,'' \emph{IEEE Control Systems Letters},
  vol.~5, no.~3, pp. 911 -- 916, 2021.

\bibitem{IEEE2008}
``{IEEE Standard for a Precision Clock Synchronization Protocol for Networked
  Measurement and Control Systems},'' \emph{IEEE Std 1588-2008 (Revision of
  IEEE Std 1588-2002)}, pp. 1--300, 2008.

\end{thebibliography}

\end{document}